# Pb-doped p-type Bi$_2$Se$_3$ thin films via interfacial engineering

*Jisoo Moon[†], Zengle Huang[†], Weida Wu[†], Seongshik Oh[†*]*

[†]Department of Physics and Astronomy, Rutgers, The State University of New Jersey, Piscataway, New Jersey 08854, United States.

*Correspondence should be addressed to ohsean@physics.rutgers.edu and +1 (848) 445-8754 (S.O.)

ABSTRACT: Due to high density of native defects, the prototypical topological insulator (TI), Bi$_2$Se$_3$, is naturally n-type. Although Bi$_2$Se$_3$ can be converted into p-type by substituting 2+ ions for Bi, only light elements such as Ca have been so far effective as the compensation dopant. Considering that strong spin-orbit coupling (SOC) is essential for the topological surface states, a light element is undesirable as a dopant, because it weakens the strength of SOC. In this sense, Pb, which is the heaviest 2+ ion, located right next to Bi in the periodic table, is the most ideal p-type dopant for Bi$_2$Se$_3$. However, Pb-doping has so far failed to achieve p-type Bi$_2$Se$_3$ not only in thin films but also in bulk crystals. Here, by utilizing an interface engineering scheme, we have achieved the first Pb-doped p-type Bi$_2$Se$_3$ thin films. Furthermore, at heavy Pb-doping, the mobility turns out to be substantially higher than that of Ca-doped samples, indicating that Pb is a less disruptive dopant than Ca. With this SOC-preserving counter-doping scheme, it is now possible to fabricate Bi$_2$Se$_3$ samples with tunable Fermi levels without compromising their topological properties.

KEYWORDS. Topological insulator, Bi$_2$Se$_3$, Doping, Interface, Spin-orbit coupling

$Bi_2Se_3$ is a three dimensional topological insulator (TI) with a large bulk band gap, ~0.3 eV, and topologically-protected surface states.[1–4] Ideally, its Fermi level ($E_F$) should be at the charge neutral point (so-called Dirac point) of the surface states within the bulk band gap. However, due to high density of n-type native defects, the Fermi levels of common $Bi_2Se_3$ are always far from the Dirac point and frequently above the bulk conduction band minimum, making the bulk metallic.[5–10] In order to fully utilize all the intriguing properties of the topological surface states, it is essential to find a way to tune the Fermi level, not only toward the Dirac point but also into the p-regime as well. In case of $Bi_2Se_3$, substituting a 2+ ion for $Bi^{3+}$ is one of the most obvious ways to reduce the n-type carriers and move toward the p-regime. As expected, soon after the confirmation of $Bi_2Se_3$ being a TI, several elements such as $Ca^{11}$, $Mn^{12}$ and $Mg^{13}$ have been found to do the job and successfully converted otherwise n-type $Bi_2Se_3$ bulk crystals into p-type. In addition, the use of Ca dopant has allowed $Bi_2Se_3$ thin films to turn into p-type not only in a relatively thick regime (~200 nm) via a complex process[14] but also in the thinnest topological regime (6 QLs).[15] However, considering that Ca is much lighter than Bi and that spin-orbit-coupling (SOC) strength, which grows fast (to the 4[th] power in a hydrogen-like atom) with increasing atomic numbers, is an essential parameter for topological surface states,[16–18] substituting light elements for Bi could have an undesirable side effect of compromising the topological properties. In this sense, a heavy 2+ ion will be a much more desirable compensation dopant than such a light element as Ca. In particular, Pb, which is located right left of Bi in the periodic table should be the most ideal p-type dopant for Bi. However, Pb-counter-doping turns out to be ineffective in converting n-type $Bi_2Se_3$ to p-type, not only in thin films but also in bulk crystals.[19,20]

Recently, however, it was found that utilization of carefully-designed buffer and capping layers can help boost the effectiveness of counter-dopants. This was first demonstrated in Ca-doping of $Bi_2Se_3$ thin films. Despite well-established p-type Ca-doped $Bi_2Se_3$ bulk crystals as discussed above, Ca-doped $Bi_2Se_3$ thin films have long failed to reach the p-regime, and only last year, the problem was solved with an interfacial engineering scheme.[15] Then, more recently, another interfacial engineering scheme has enabled Ti-doping to convert otherwise p-type $Sb_2Te_3$ thin films into n-type.[21] Ti is a potential n-type dopant for $Sb^{3+}$ ion in $Sb_2Te_3$. However, it had long failed to reach the n-regime of $Sb_2Te_3$ until a proper interfacial engineering scheme was developed recently. Here, we show that the interfacial engineering scheme now allows Pb-counter-doping to implement p-type $Bi_2Se_3$ for the first time in any $Bi_2Se_3$ samples, either thin films or bulk crystals.

Films were grown on $10 \times 10$ mm$^2$ c-plane (0001) sapphire ($Al_2O_3$) substrates using molecular beam epitaxy (MBE) in a custom-built UHV system with base pressure below $5 \times 10^{-10}$ torr. The individual sources of high purity (99.999%) Bi, In, Se, and Pb were evaporated from standard effusion cells during the film growth. Film thickness and Pb-doping levels were determined with Rutherford backscattering spectroscopy and quartz crystal microbalance (QCM). Se flux was maintained at least 10 times higher than those of Bi and In to minimize Se vacancies. DC transport measurements were carried out with the standard Van der Pauw geometry using a cryogenic resistive electromagnet system whose base temperature is 6 K. The Pb-doped $Bi_2Se_3$ films were grown on $(Bi_{0.5}In_{0.5})_2Se_3$ (BIS in short) buffer layers composed of 20 QLs of $(Bi_{0.5}In_{0.5})_2Se_3$ and $In_2Se_3$, respectively, as shown in Figure 1. Since $In_2Se_3$ develops multiple phases when grown directly on the sapphire substrates, 3 QL $Bi_2Se_3$ is grown first at 135 °C as a seed layer, followed by 20 QL $In_2Se_3$ grown at 300 °C and annealing to 600 °C, which evaporates out the $Bi_2Se_3$ seed layer. Then, we cool it down to 275 °C and deposit 20 QLs of $(Bi_{0.5}In_{0.5})_2Se_3$ and Pb-doped $Bi_2Se_3$ in sequence. Amorphous Se, approximately 100 nm thick, is deposited at room temperature as a capping layer. More detailed information for the BIS buffer layer growth can be found in Ref. 22.

Figure 2 shows doping-dependent Hall effect data of the 20 QL Pb-doped $Bi_2Se_3$ films. As the doping concentration increases, the hole dopants compensate for the n-type carriers in the $Bi_2Se_3$ film, leading to increased negative slopes of the Hall effect curve, implying reduced n-type sheet carrier densities ($n_{2D}$s), until the doping level reaches 0.05%. As the Pb-doping increases further to 0.072%, the negative slope starts to decrease, indicating that the films starts to have p-type carriers in addition to the majority n-type carriers. A decreasing slope of a Hall effect curve in $Bi_2Se_3$ as the Fermi level approaches the charge neutral point is a commonly observed behavior both with compensation doping and with gating.[15,23,24] At 0.1%, the Hall curve becomes almost flat, implying that there exist significant level of electron-hole puddles, spatially varying energy levels near the Dirac point[25] which is analogous to similar features in graphene.[26–28] At 0.2%, the slope increases to a clear positive value, indicating that the Fermi level has now exited the electron-hole puddle regime and clearly entered the p-regime. Up to 1%, the carrier type remains p-type. However, at much higher doping levels, the carrier type changes back to n-type as shown in the 10% data, and the (n-type) $n_{2D}$ becomes extremely high at 20%. Such a carrier-type reversal behavior at high doping concentrations was previously observed in the Ca-doped $Bi_2Se_3$ film study,[15] and indicates that there is a solubility limit for any compensation dopant. Beyond a solubility limit, the dopants start to introduce extra defects going beyond simple Bi substitution, and it is well known that almost all defects in $Bi_2Se_3$ behave as n-type dopants.[5,6]

The $n_{2D}$s and mobilities ($\mu$s) extracted from Figure 2 are shown in Figure 3, where $n_{2D} = 1/(e \cdot \text{slope})$ and $\mu = 1/(R_{\text{sheet}} \cdot n_{2D} \cdot e)$. Here the data for 0.1% and 20% are not included because they give orders of magnitude higher - nominal for the former and actual for the latter - $n_{2D}$s than the others. Figure 3(a) shows what we have quantitatively discussed above with Figure 2 on how the $n_{2D}$s and their signs change with Pb-doping: n-type up to 0.05%, n-p mixing for 0.072% and 0.1%, p-type up to 1.0%, and finally back to n-type for even higher Pb-doping. The mobility data in Figure 3(b) provide further information. The most notable feature is that the $\mu$s of p-type samples are clearly lower than those of n-type samples, which is consistent with the previous Ca-doped $Bi_2Se_3$ study.[15] The very fact that the 10% n-type sample exhibits a higher mobility than the sub 1% p-type samples strongly suggests that the higher mobility of n-type than p-type samples should be an intrinsic band structure effect rather than an extrinsic dopant-scattering effect. This is also consistent with the known band structure of $Bi_2Se_3$: the surface band structure is much sharper above the Dirac point than below,[3,9,29–31] thus higher Fermi velocity (or/and lower effective mass) and higher mobility for n-type than p-type. Furthermore, the proximity of the bulk valence band to the Dirac point also leads to more scattering channels, thus lower mobility, for the p-type carriers.

In Figure 4, we present scanning tunneling microscopy (STM) images of $Bi_2Se_3$ films at two different Pb-doping levels (0.6% and 6%) to probe the level of disorder introduced by the doping process. While the 0.6% sample in Figure 4(a) shows the prototypical triangular-shaped step terraces,[32] the 6% sample in Figure 4(b) is filled with more irregular morphology. This clearly indicates that high level of Pb-doping in $Bi_2Se_3$ is accompanied by disorder, as expected from the transport data above. The zoomed-in images of the 0.6% sample in Figure 4(c-d) exhibit almost uniformly-distributed Pb-dopants. Interestingly, the counted Pb-dopant density, 5.4%, is almost 10 times higher than the nominal concentration of 0.6% estimated by QCM. This implies that there is a strong tendency for Pb-dopants to diffuse out of the bulk of $Bi_2Se_3$ and stay on the interfaces. Then the actual doping levels of the bulk and interfaces are expected to be substantially different (the former being lower and the latter higher) than the nominal doping level.

Finally, we compare highly Pb- and Ca-doped $Bi_2Se_3$ thin films in Table 1. For small levels (< 1%), we do not find noticeable differences between the two doping schemes, but at high doping concentrations (> 1%), which inevitably introduces significant disorders, the difference becomes noticeable. Table 1 shows that the impact of disorder is less disruptive with Pb than with Ca doping at high doping. First of all, even with much higher Pb doping, the $n_{2D}$ is slightly lower ($1.03 \times 10^{13}$ vs. $1.50 \times 10^{13}$ $cm^{-2}$) than the Ca doping, implying that Pb doping entails fewer defects. Second and more noticeably, the $\mu$ is substantially higher (949 vs. 217 $cm^2$/V·s) with Pb doping, implying that defects from the Pb-dopant introduce less scattering centers than those from Ca. Although the lower $n_{2D}$ (fewer defects) may be related to the strong tendency for Pb dopants to diffuse out of the bulk, the higher mobility is likely due to the much stronger SOC strength of Pb than Ca. Strong SOC is a necessary condition for the absence of backscattering for topological surface states, so light dopant like Ca is much more likely to induce backscattering than Pb. Further studies will be needed to fully understand the role of different SOC strengths in compensation-doped topological insulators.

In conclusion, we have achieved the first Pb-doped p-type $Bi_2Se_3$ system. The key is utilization of the BIS buffer layer in combination with a protective capping layer. It is particularly notable that the interface engineering scheme enabled access to a doping regime that cannot be accessed in bulk crystals with the same dopant. This is in stark contrast with the Ca-doped $Bi_2Se_3$ system, for which p-type bulk crystals have long existed before p-type thin films became available with an interface engineering scheme. When compared with Ca doping, Pb-doping has a clear advantage of preserving the strong SOC of the TI system: the much higher mobility of a heavily Pb-doped $Bi_2Se_3$ than a Ca-doped one seems to support this idea. The ability to control the Femi level of a TI system without compromising their topological properties will provide new opportunities for engineering topological materials.

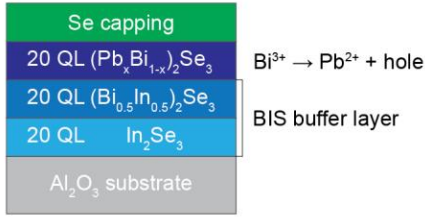

**Figure 1.** Schematic layer structure of the 20 QL Pb-doped $Bi_2Se_3$ films. The Pb-doped $Bi_2Se_3$ is grown on the BIS buffer layer and followed by a Se capping layer. The dopants of $Pb^{2+}$ introduce holes by replacing the $Bi^{3+}$ atoms in the $Bi_2Se_3$ system. Pb-doping in $Bi_2Se_3$ films is feasible only on top of the BIS buffer layer scheme in combination with proper capping layers.

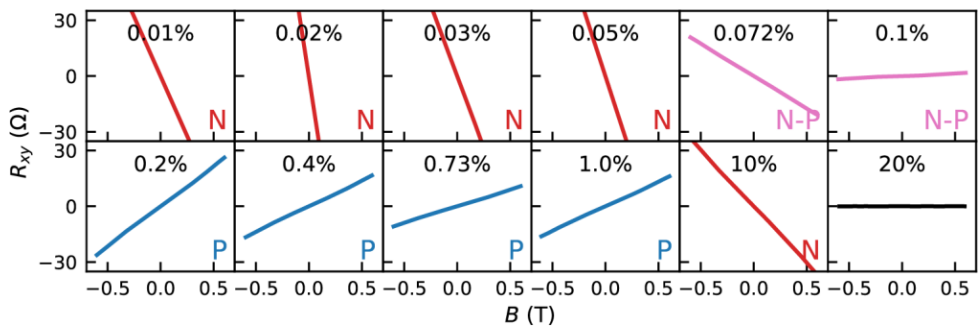

**Figure 2.** Doping-dependent Hall effect data for the 20 QLs of Pb-doped $Bi_2Se_3$ films at 6 K. N-type (negative slope), n-p mixing, and p-type (positive slope) curves are colored red, pink, and blue, respectively. The n-regime ranges up to 0.05%. There exists an n-p mixed regime around 0.072 ~ 0.1% between n- and p-type. The 20 QL Pb-doped $Bi_2Se_3$ film becomes p-type from 0.2% to 1.0%. The carrier type turns back to n-type above 1% as shown in the 10% data. The $n_{2D}$ becomes very high at 20%, thus it is difficult to recognize the carrier type, which is indicated by the nearly flat Hall effect data and colored as black.

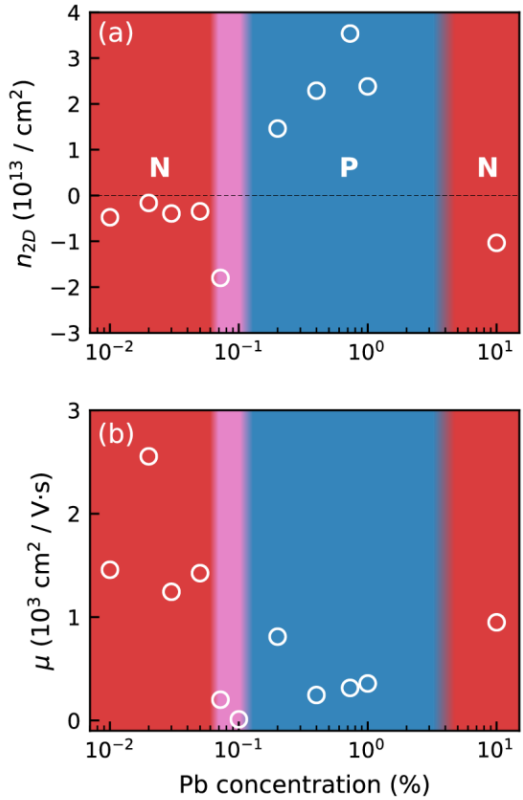

**Figure 3.** Transport properties of the 20 QL Pb-doped $Bi_2Se_3$ films. (a) Sheet carrier density and (b) mobility as functions of the Pb doping concentration. Negative and positive $n_{2D}$ data indicate n- and p-type carriers in (a). The n-, n-p, and p-regime are colored by red, pink, and blue, respectively. As the doping concentration increases, the $Bi_2Se_3$ film undergoes a carrier type transition from n to p passing through a n-p mixing regime. At high concentrations, the film becomes n-type due to disorder beyond the solubility limit of Pb in $Bi_2Se_3$, accompanied by a higher (than p-type) mobility shown in (b).

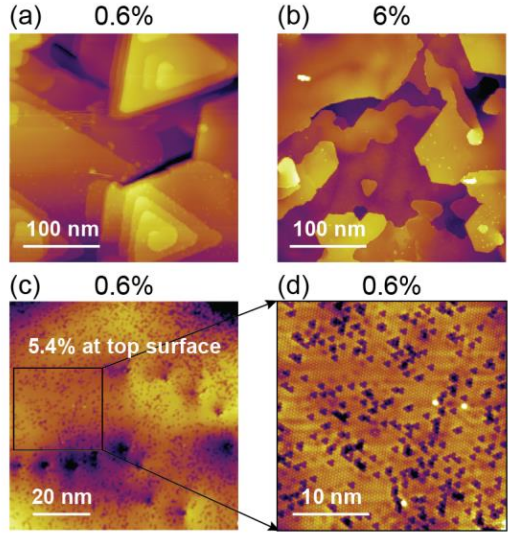

**Figure 4.** STM images for morphology comparison and doping level investigation at the top surface. (a-b) Top surface topography images of 20 QL Pb-doped $Bi_2Se_3$ films with (a) 0.6% and (b) 6% of Pb-doping. (c) Zoom-in image in which Pb dopants are visible. (d) Zoom-in image of the area marked by a square in (c). The investigation shows that the concentration of Pb on the top surface is estimated as 5.4%, which is nearly 10 times higher than the nominal concentration of 0.6%.

|  | $n_{2D}$ ( / $cm^2$ ) | $\mu$ ( $cm^2$ / V·s ) |
|---|---|---|
| 10% of Pb at 6 K | $1.03 \times 10^{13}$ | 949 |
| 4.4% of Ca at 2 K | $1.50 \times 10^{13}$ | 217 |

**Table 1.** Comparison of transport properties in 20 QLs of highly Pb- and Ca-doped $Bi_2Se_3$. The data of Ca-doped $Bi_2Se_3$ are taken from Ref. 15. The mobility of Pb-doped film is substantially higher than that of Ca-doped one, which means that Pb-doping is less disruptive than Ca-doping in $Bi_2Se_3$.